# Optimal Transmit Power Allocation for MIMO Two-Way Cognitive Relay Networks with Multiple Relays

Ahmad Alsharoa, *Member, IEEE*, Hakim Ghazzai, *Member, IEEE*, and Mohamed-Slim Alouini, *Fellow, IEEE*

*Abstract*—In this letter, we consider a multiple-input multiple-output two-way cognitive radio system under a spectrum sharing scenario, where primary and secondary users operate on the same frequency band. The secondary terminals aims to exchange different messages with each other using multiple relays where each relay employs an amplify-and-forward strategy. The main objective of our work is to maximize the secondary sum rate allowed to share the spectrum with the primary users by respecting a primary user tolerated interference threshold. In this context, we derive an analytical expression of the optimal power allocated to each antenna of the terminals. We then discuss the impact of some system parameters on the performance in the numerical result section.

## I. INTRODUCTION

**N**OWADAYS improvement of a spectrum usage and data rate have gained enormous importance and attention from researchers in the fields of wireless communications. Several solutions including Cognitive Radio (CR), cooperative communication, and Multi-Input Multi-Output (MIMO) have been proposed. The ideas have centered around incorporating two or more of these schemes together to solve the spectral limitation and the high data rate demand issues.

During the last decade, CR has been introduced as one of the promising solutions to solve the spectrum utilization problems [1]. CR spectrum sharing allows Secondary Users (SUs) to utilize the frequency band allocated by Primary Users (PUs). As such and in order to protect the PUs, the transmit power of SUs should be controlled to satisfy a certain primary Quality of Service (QoS) measured by an interference threshold [2].

On the other hand, cooperative communications and particularly Two-Way Relaying (TWR) have been proposed recently as a smart solution to increase the overall system throughput and enhance the network coverage area [3]. In conventional TWR, exchanging different messages between two terminals takes place in two time slots only. In the first time slot, the terminals transmit their signals simultaneously to the relays.

The authors are with the Electrical Engineering Program, Computer, Electrical and Mathematical Sciences and Engineering (CEMSE) Division, King Abdullah University of Science and Technology (KAUST), Thuwal, Makkah Province, Saudi Arabia. (E-mails: {ahmad.sharoa, hakim.ghazzai, slim.alouini}@kaust.edu.sa).

The work of Mohamed-Slim Alouini was made possible by NPRP grant #5 − 250 − 2 − 087 from the Qatar National Research Fund (a member of Qatar Foundation). The statements made herein are solely the responsibility of the authors.

Subsequently, in the second time slot, the relays broadcast the signal to the terminals [3]. Several studies have been proposed to analyze CR with cooperative relaying technique using Amplify-and-Forward (AF) strategy [4], [5], where the AF strategy causes less delays and requires lower hardware complexity compare to other relay strategies. The authors in [6] investigate the optimal power allocation at the terminals and single relay selection for TWR-CR networks under single antenna scenario.

An overview of TWR under MIMO scenario has been presented in [7]. Most existing models of MIMO TWR are proposed for single relay case [8], [9]. Indeed, more efficient usage of the spectrum can be achieved by combining CR, cooperative communication, and MIMO antennas together. However, to the best knowledge of the authors, the multiple relay problem in MIMO TWR-CR networks has not been discussed so far as it is the case for non-cognitive case.

In this letter, a multiple relay scheme for MIMO TWR-CR networks with AF strategy is investigated. The main contributions of this paper can be summarized as follows: (i) Formulate an optimization problem for multiple relay MIMO TWR-CR networks with AF strategy that aims to maximize the secondary sum rate by taking into account the power budget of the system and the interference level tolerated by the PUs, (ii) Derive the optimal solution expression of the power allocation problem at each antenna of the terminals, and (iii) Analyze the impact of the system parameters on the scheme performance.

*Notations*: The superscript $(.)^T$ and $(.)^H$ denote the transpose operator and the hermitian operator, respectively. $\mathbb{C}$ and $\mathbb{R}$ refer to the field of complex and real numbers, respectively. $\mathbb{E}(.)$ and $\text{Tr}(.)$ denote the expectation and the trace operator, respectively.

## II. SYSTEM MODEL

In this letter, cognitive system consisting of one PU and a cognitive network is considered. As illustrated in Fig.1, the cognitive network is constituted of two cognitive transceiver terminals $T_1$ and $T_2$, and $L$ cognitive half-duplex relays. The PU, $T_1$, $T_2$, and the $i^{th}$ relay are equipped by $M_{PU}, M_{T_1}, M_{T_2}$ and $M_{R_i}$ antennas, respectively. A non-line of sight link between $T_1$ and $T_2$ is also considered. In this work, we assume that the PU and SUs access the spectrum simultaneously. In order to protect the PU, the average received interference power due to the secondary nodes should be below a certain interfernce threshold denoted $I_{th}$ [2]. Without loss of generality, all the noise variances are assumed to be equal to $N_0$.

During the first time slot, $T_1$ and $T_2$ transmit their signals $\tilde{x}_1$ and $\tilde{x}_2$ to the relays simultaneously, with a power denoted $\boldsymbol{P}_1 = [P_1^1, ..., P_1^{M_{T_1}}]$ and $\boldsymbol{P}_2 = [P_2^1, ..., P_2^{M_{T_2}}]$, respectively. In the second time slot, the relays transmit the amplified signals to the terminals, with a power denoted $\boldsymbol{P}_{r_i} = [P_{r_i}^1, ..., P_{r_i}^{M_{R_i}}]$, where $i = 1, ..., L$.

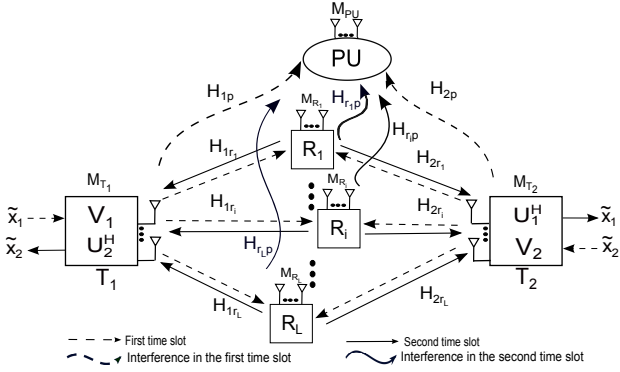

Fig. 1 MIMO TWR-CR system model.

Let us define $\bar{P}_t, \bar{P}_r, \boldsymbol{H}_{1r_i} \in \mathbb{C}^{M_{R_i} \times M_{T_1}}, \boldsymbol{H}_{2r_i} \in \mathbb{C}^{M_{R_i} \times M_{T_2}}, \boldsymbol{H}_{r_i p} \in \mathbb{C}^{M_{R_i} \times M_{PU}}, \boldsymbol{H}_{1p} \in \mathbb{C}^{M_{T_1} \times M_{PU}}$, and $\boldsymbol{H}_{2p} \in \mathbb{C}^{M_{T_2} \times M_{PU}}$ as the peak power at the transceiver terminals, peak power at each relay, the complex channel mapping matrix between $T_1$ and the $i^{th}$ relay, the complex channel mapping matrix between $T_2$ and the $i^{th}$ relay, the complex channel mapping matrix between the $i^{th}$ relay and the PU, the complex channel mapping matrix between $T_1$ and the PU, and the complex channel mapping matrix between $T_2$ and the PU, respectively. All the channel gains adopted in our framework are assumed to be constant during the coherence time with elements $h_{ab}^{xy}$ representing the fading coefficients between transmit antenna $y$ at node $a$ and receive antenna $x$ at node $b$. We assume that full Channel State Information (CSI) are available at both terminals. This is not a very benign assumption as feedback CSI from PU to SUs is adopted in many cognitive works [10].

Let $V_m$ and $U_m$, where $m = 1, 2$, two unitary precoder and decoder matrices, respectively. In the first time slot, $T_1$ and $T_2$ employ the precoder matrices $\boldsymbol{V}_1$ and $\boldsymbol{V}_2$, respectively, such as: $\boldsymbol{x}_1 = \boldsymbol{V}_1 \tilde{\boldsymbol{x}}_1$ and $\boldsymbol{x}_2 = \boldsymbol{V}_2 \tilde{\boldsymbol{x}}_2$, where $\boldsymbol{x}_m$ is the transmitted signal after being precoded by $T_m$. Subsequently, during the second time slot, $T_1$ and $T_2$ employ the decoder matrices $\boldsymbol{U}_2$ and $\boldsymbol{U}_1$, respectively, such as: $\boldsymbol{r}_2^{(T_1)} = \boldsymbol{U}_2^H \boldsymbol{y}_2^{(T_1)}$ and $\boldsymbol{r}_1^{(T_2)} = \boldsymbol{U}_1^H \boldsymbol{y}_1^{(T_2)}$, where $\boldsymbol{y}_2^{(T_1)}$ and $\boldsymbol{r}_2^{(T_1)}$ are the received signals at $T_1$ before and after decoding, respectively, while $\boldsymbol{y}_1^{(T_2)}$ and $\boldsymbol{r}_1^{(T_2)}$ are the received signals at $T_2$ before and after decoding, respectively. It is also assumed that $\mathbb{E}\left(||\boldsymbol{x}_m||^2\right) = \mathrm{Tr}\left(\boldsymbol{x}_m \boldsymbol{x}_m^H\right) \leq \bar{P}_t$. The choice of $\boldsymbol{V}_m$ and $\boldsymbol{U}_m$ will be defined later.

## III. PROBLEM FORMULATION AND OPTIMAL POWER ALLOCATION

In this section, we formulate the optimization problem that maximizes the secondary sum rate for MIMO TWR-CR networks without affecting the QoS of the PU. For simplicity, we assume that all the relays are equipped with the same number of antennas, i.e., $M_{R_i} = M_R, \forall i = 1, ..., L$.

In the first time slot, the complex baseband received signal at the $i^{th}$ relay is given by

$$\boldsymbol{y}_{r_i} = \boldsymbol{H}_{1r_i} \boldsymbol{x}_1 + \boldsymbol{H}_{2r_i} \boldsymbol{x}_2 + \boldsymbol{n}_{r_i}, \quad (1)$$

where $\boldsymbol{n}_{r_i}$ is the additive Gaussian noise vector at the $i^{th}$ relay and $\boldsymbol{x}_m$ is the transmitted signal after precoding by terminal $T_m$. During the second time slot, each relay amplifies $\boldsymbol{y}_{r_i}$ by multiplying it by a diagonal matrix $\boldsymbol{W}_i \in \mathbb{R}^{M_R \times M_R}$ (containing the amplification factor $w_i^k$ at each antenna $k$ of the $i^{th}$ relay) and broadcasts it to the terminals $T_1$ and $T_2$. $\boldsymbol{W}_i, \forall i = 1, \ldots, L$, are diagonal matrices as, in our case, we are using blind amplification relays [11]. Finally, the received signals $\boldsymbol{y}_2^{(T_1)}$ and $\boldsymbol{y}_1^{(T_2)}$ at $T_1$ and $T_2$ are given, respectively as

$$\begin{cases} \boldsymbol{y}_2^{(T_1)} = \underbrace{\hat{\boldsymbol{A}}_2 \boldsymbol{x}_1}_{\text{Self Interference}} + \boldsymbol{A}_2 \boldsymbol{x}_2 + \boldsymbol{z}_2^{(T_1)}, \\ \boldsymbol{y}_1^{(T_2)} = \boldsymbol{A}_1 \boldsymbol{x}_1 + \underbrace{\hat{\boldsymbol{A}}_1 \boldsymbol{x}_2}_{\text{Self Interference}} + \boldsymbol{z}_1^{(T_2)}, \end{cases} \quad (2)$$

where $\boldsymbol{A}_2 = \sum_{i=1}^{L} \boldsymbol{H}_{1r_i}^T \boldsymbol{W}_i \boldsymbol{H}_{2r_i}, \boldsymbol{A}_1 = \sum_{i=1}^{L} \boldsymbol{H}_{2r_i}^T \boldsymbol{W}_i \boldsymbol{H}_{1r_i}, \hat{\boldsymbol{A}}_2 = \sum_{i=1}^{L} \boldsymbol{H}_{1r_i}^T \boldsymbol{W}_i \boldsymbol{H}_{1r_i}, \hat{\boldsymbol{A}}_1 = \sum_{i=1}^{L} \boldsymbol{H}_{2r_i}^T \boldsymbol{W}_i \boldsymbol{H}_{2r_i}, \boldsymbol{z}_2^{(T_1)} = \sum_{i=1}^{L} \left(\boldsymbol{H}_{1r_i}^T \boldsymbol{W}_i \boldsymbol{n}_{r_i}\right) + \boldsymbol{n}_1, \boldsymbol{z}_1^{(T_2)} = \sum_{i=1}^{L} \left(\boldsymbol{H}_{2r_i}^T \boldsymbol{W}_i \boldsymbol{n}_{r_i}\right) + \boldsymbol{n}_2$, and $\boldsymbol{n}_m$ is the additive Gaussian noise vector at $T_m$, where $m = 1, 2$. By using the knowledge of the side information and channel reciprocity, the terminals can remove the self interference by eliminating their own signals (i.e., $\boldsymbol{x}_1$ for $T_1$ and $\boldsymbol{x}_2$ for $T_2$). Then, we propose to define the precoding and decoding matrices based on the Singular Value Decomposition (SVD) of the matrices $\boldsymbol{A}_m = \boldsymbol{U}_m \boldsymbol{\Sigma}_m \boldsymbol{V}_m^H, m = 1, 2$.

The amplification factor at the $k^{th}$ antenna of the $i^{th}$ relay can be expressed as

$$|w_i^k|^2 = \frac{P_{r_i}^k}{\sum_{z=1}^{M_{T_1}} P_1^z |h_{1r_i}^{kz}|^2 + \sum_{z=1}^{M_{T_2}} P_2^z |h_{2r_i}^{kz}|^2 + N_0}, \quad (3)$$

where $P_{r_i}^k$ denotes as the power at the $k^{th}$ antenna of the $i^{th}$ relay. Thus, the sum rate of the MIMO TWR after SVD can be written as

$$R = \frac{1}{2} \sum_{u=1}^{M_{min}} \log_2 \left(1 + \frac{\sigma_{2u}^2 P_2^u}{N_2}\right) + \frac{1}{2} \sum_{v=1}^{M_{min}} \log_2 \left(1 + \frac{\sigma_{1v}^2 P_1^v}{N_1}\right), \quad (4)$$

where $\begin{cases} N_1 = N_0 \left(1 + \sum_{i=1}^{L} \sum_{v=1}^{M_{T_1}} \sum_{k=1}^{M_R} |w_i^k h_{1r_i}^{kv}|^2\right), \\ N_2 = N_0 \left(1 + \sum_{i=1}^{L} \sum_{u=1}^{M_{T_2}} \sum_{k=1}^{M_R} |w_i^k h_{2r_i}^{ku}|^2\right), \end{cases} \quad (5)$

$M_{min} = \min(M_{T_1}, M_{T_2})$, and $\sigma_{mq}$ is the $q^{th}$ diagonal element of the matrix $\boldsymbol{\Sigma}_m$. Thus, the sum rate maximization problem of MIMO TWR-CR with multiple relays can now be formulated as expressed in (6). Note that this optimization problem deals with the terminal side only by maximizing the secondary sum rate without any control on relay parameters. Adjusting relay and terminal powers simultaneously is a possible interesting future extension of this work.

$$\underset{\boldsymbol{P}_1,\boldsymbol{P}_2}{\text{maximize}} \quad R \tag{6}$$

subject to
(C1: Peak power constraint at the terminals):
$$0 \leq \sum_{v=1}^{M_{T_1}} P_1^v \leq \bar{P}_t, \quad 0 \leq \sum_{u=1}^{M_{T_2}} P_2^u \leq \bar{P}_t, \tag{7}$$

(C2: Peak power constraint at the relays):
$$0 \leq \sum_{k=1}^{M_R}\left(\sum_{v=1}^{M_{T_1}} P_1^v|h_{1r_i}^{kv}|^2 + \sum_{u=1}^{M_{T_2}} P_2^u|h_{2r_i}^{ku}|^2 + N_0\right)|w_i^k|^2 \leq \bar{P}_r,$$
$$\forall i = 1,...,L, \tag{8}$$

(C3: Interference constraint in the first time slot):
$$\sum_{v=1}^{M_{T_1}}\sum_{j=1}^{M_{PU}} P_1^v|h_{1p}^{jv}|^2 + \sum_{u=1}^{M_{T_2}}\sum_{j=1}^{M_{PU}} P_2^u|h_{2p}^{ju}|^2 \leq I_{th}, \tag{9}$$

(C4: Interference constraint in the second time slot):
$$\sum_{i=1}^{L}\sum_{j=1}^{M_{PU}}\sum_{k=1}^{M_R}\left(\sum_{v=1}^{M_{T_1}} P_1^v|h_{1r_i}^{kv}|^2 + \right.$$
$$\left.\sum_{u=1}^{M_{T_2}} P_2^u|h_{2r_i}^{ku}|^2 + N_0\right)|w_i^k|^2|h_{r_ip}^{jk}|^2 \leq I_{th}. \tag{10}$$

To solve this problem, we propose to use the Lagrangian method [12]. The Lagrangian function $\mathcal{L}$ is derived in (11) where $\boldsymbol{\lambda}$ is the vector that contains all the Lagrangian multipliers of the system. $\lambda_1, \lambda_2,$ and $\lambda_{r_i}$ represent the Lagrangian multipliers related to the peak power at $T_1$, $T_2$, and $i^{th}$ relay, while $\lambda_{th_1}$ and $\lambda_{th_2}$ represent the Lagrangian multipliers related to the first and second time slot interference constraints, respectively. By taking the derivative of $\mathcal{L}$ with respect to the $P_m^q$ where $q = 1 : M_{T_m}, m = 1, 2$, we find the optimal transmit power allocated to the $q^{th}$ antenna at the terminals that maximizes the Lagrangian function and the sum rate. Its expression is given in (12), where $(x)^+$ denotes the maximum between $x$ and zero. Therefore, we can solve our convex optimization problem by exploiting its strong duality [12].

$$\min_{\boldsymbol{\lambda} \geq 0} \max_{\boldsymbol{P}_1 \geq 0, \boldsymbol{P}_2 \geq 0} \mathcal{L}(\boldsymbol{\lambda}, \boldsymbol{P}_1, \boldsymbol{P}_2). \tag{13}$$

We can employ the subgradient method or other heuristic approaches to find the optimal Lagrangian multipliers of this problem [13].

## IV. SIMULATION RESULTS

In this section, numerical results are provided for identically distributed Rayleigh fading channels. All the communication terminals of the system are equipped with the same number of antennas (i.e., $M_{T_1} = M_{T_2} = M_{PU} = M_T$). The noise variance $N_0$ is assumed to be equal to $10^{-4}$.

Fig.2 plots the achieved secondary sum rate versus terminals peak power ($\bar{P}_t$) for different values of $I_{th} = \{10, 20\}$ dBm and different values of $L = \{2, 6\}$ with fixed entries of $\boldsymbol{W}$ as a matrix and $\bar{P}_r$. We can notice that the achievable rate is improving with the increase of $\bar{P}_t$ up to a certain value, due to the fact that starting from this value of $\bar{P}_t$ the system can no more supply the secondary terminals with their full power budget because the system has to respect constraints (9) and (10). For this reason, we have introduced a MIMO scheme in order to get more degrees of freedom by increasing the number of antennas and as a result enhance the secondary sum rate. For instance, for $L = 2$ and $I_{th} = 20$, we were able to improve the achievable secondary sum rate by $180\%$ by going from around 5 bits/s/Hz to 14 bits/s/Hz at high values of $\bar{P}_t$ with the usage of four antennas instead of a single one. This figure shows also the validity of the derived analytical solution expressed in (12) which is compared to the optimal solution obtained using simulations.

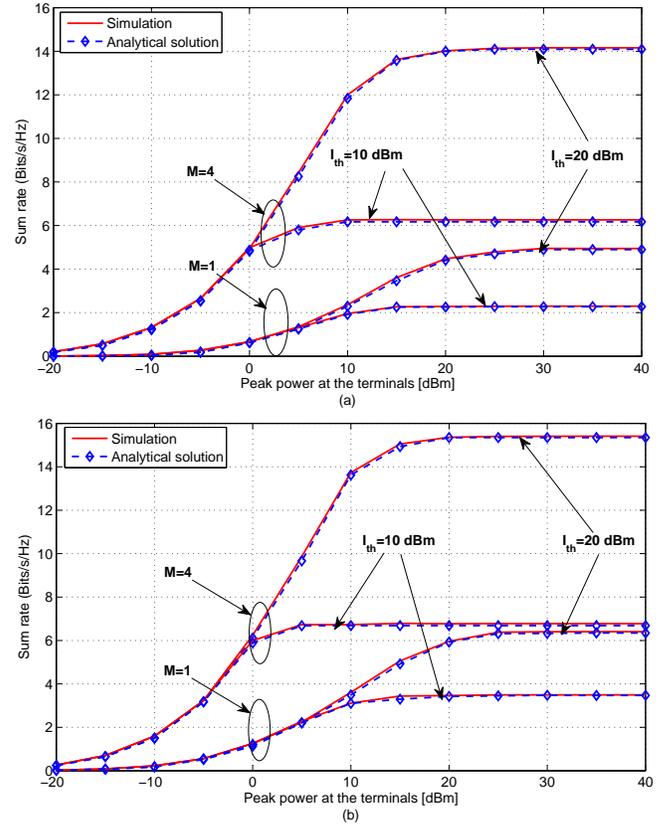

Fig. 2. Achieved secondary sum rate versus peak terminal power constraint with $w = .2, M_T = M_R = M$ and $\bar{P}_r = 20$ dBm, for (a) $L = 2$, (b) $L = 6$.

In the following, we aim to investigate the impact of the relay amplification factors on the system performance. We assume that the secondary network is constituted by similar relays having same amplification factors (i.e., $\boldsymbol{W}_i = w\boldsymbol{I}, \forall i = 1, ..., L$). The initial system parameters are given as the following: $\bar{P}_t = 20$ dBm, $\bar{P}_r = 10$ dBm, $I_{th} = 20$ dBm, $M_T = 2$, $M_R = 2$, and $L = 4$. At each subfigure of Fig.3, where we plot the achievable sum rate versus the amplification factor ($w$), we vary one of the previously cited parameters by keeping all the others fixed. It is clear that increasing one of these parameters will trivially enhance the sum rate.

$$\mathcal{L}(\boldsymbol{\lambda},\boldsymbol{P}_1,\boldsymbol{P}_2) = \tfrac{1}{2}\sum_{u=1}^{M_{min}}\log_2\left(1+\tfrac{\sigma_{2u}^2 P_2^u}{N_2}\right) + \tfrac{1}{2}\sum_{v=1}^{M_{min}}\log_2\left(1+\tfrac{\sigma_{1v}^2 P_1^v}{N_1}\right) - \lambda_1\left(\sum_{v=1}^{M_{T_1}} P_1^v - \bar{P}_t\right) - \lambda_2\left(\sum_{u=1}^{M_{T_2}} P_2^u - \bar{P}_t\right) -$$
$$\sum_{i=1}^{L}\lambda_{r_i}\left(\sum_{k=1}^{M_R}\left(\sum_{v=1}^{M_{T_1}} P_1^v|h_{1r_i}^{kv}|^2 + \sum_{u=1}^{M_{T_2}}|h_{2r_i}^{ku}|^2 + N_0\right)|w_i^k|^2 - \bar{P}_r\right) - \lambda_{th_1}\left(\sum_{v=1}^{M_{T_1}}\sum_{j=1}^{M_{PU}} P_1^v|h_{1p}^{jv}|^2 + \sum_{u=1}^{M_{T_2}}\sum_{j=1}^{M_{PU}} P_2^u|h_{2p}^{ju}|^2 - I_{th}\right) \quad (11)$$
$$-\lambda_{th_2}\left(\sum_{i=1}^{L}\sum_{j=1}^{M_{PU}}\sum_{k=1}^{M_R}\left(\sum_{v=1}^{M_{T_1}} P_1^v|h_{1r_i}^{kv}|^2 + \sum_{u=1}^{M_{T_2}} P_2^u|h_{2r_i}^{ku}|^2 + N_0\right)|w_i^k|^2|h_{r_ip}^{jk}|^2 - I_{th}\right).$$

$$P_m^q = \left(\frac{1}{2\log_e 2\left[\lambda_m + \sum_{i=1}^{L}\lambda_{r_i}\sum_{k=1}^{M_R}|h_{mr_i}^{kq}w_i^k|^2 + \lambda_{th_1}\sum_{j=1}^{M_{PU}}|h_{mp}^{jq}|^2 + \lambda_{th_2}\sum_{i=1}^{L}\sum_{j=1}^{M_{PU}}\sum_{k=1}^{M_R}|h_{mr_i}^{kq}w_i^k h_{r_ip}^{jk}|^2\right]} - \frac{N_m}{\sigma_{mq}^2}\right)^+, m = \{1,2\}. \quad (12)$$

First, in Fig.3(a) we vary $\bar{P}_t$. As we increase it; the optimal value of $w$ denoted by $w_{opt}$ (corresponding to the vertical dashed lines), for which the sum rate achieves its maximum, decreases. This can be justified by the fact that, as a direct reaction to the increase of $\bar{P}_t$, $w_{opt}$ will be reduced in order to satisfy both power and interference constraints, respectively. In contrary, as shown in Fig.3(b) increasing $\bar{P}_r$ will increase $w_{opt}$ where the maximum rate is achieved as clearly shown from the relation between them in (8). However, increasing $I_{th}$ will trivially help in increasing the sum rate without affecting $w_{opt}$ since the power budget of terminals and relays remain constant. Finally, Fig.3(e) and Fig.3(f) show that when $M_R$ or $L$ increase, $w_{opt}$ decreases because the term multiplying by $w$ in (10) will increase, thus $w_{opt}$ should decrease to satisfy this constraint. One can see that, increasing $L$ affects more the sum rate than increasing $M_R$ since adding relays is more beneficial than adding antennas.

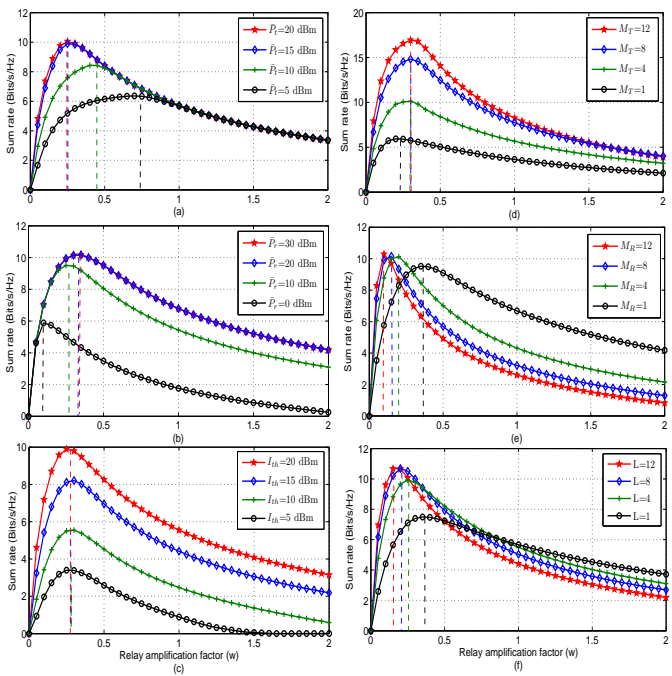

Fig. 3. Achieved sum rate versus relay amplification factor $w$.

## V. CONCLUSION

In this letter, we have derived an analytical expression of an optimal transmit power allocation scheme for MIMO TWR-CR networks. The idea of this scheme is to maximize the sum rate of the secondary networks without degrading the PU performance by imposing an interference constraints to the secondary network. Moreover, the amplification factor of the relay side is analyzed for different system parameters. In our future work, we will focus on the optimization of the sum rate by dealing with the relay amplification factors simultaneously with the transmit power allocation either by deriving an optimal solution or by employing an heuristic approach.